\documentclass[fleqn,10pt]{wlscirep}
\usepackage[utf8]{inputenc}
\usepackage[T1]{fontenc}

\usepackage{amsmath}
\usepackage{graphicx}
\usepackage[colorlinks=true,allcolors=blue]{hyperref}
\usepackage{caption}
\usepackage{subcaption}
\usepackage{tabularx}
\usepackage{svg}

\title{Twisted by the Pools: Detection of Selfish Anomalies in Proof-of-Work Mining}

\author[1,2,*]{Sheng-Nan Li}
\author[3,1,2]{Carlo Campajola}
\author[1,2]{Claudio J. Tessone}
\affil[1]{Blockchain \& Distributed Ledger Technologies, Faculty of Business, Economics and Informatics, University of Zurich, CH-8050, Switzerland}
\affil[2]{UZH Blockchain Center, Faculty of Business, Economics and Informatics, University of Zurich, CH-8050, Switzerland}
\affil[3]{DLT Science Foundation, London, United Kingdom}

\affil[*]{shengnan.li@uzh.ch}

\begin{abstract}
The core of many cryptocurrencies is the decentralised validation network operating on proof-of-work technology. In these systems, validation is done by so-called \textit{miners} who can digitally sign blocks once they solve a computationally-hard problem.  Conventional wisdom generally considers this protocol as secure and stable as miners are incentivised to follow the behaviour of the majority. However, whether some strategic mining behaviours occur in practice is still a major concern. In this paper we target this question by focusing on a security threat: a selfish mining attack in which malicious miners deviate from protocol by not immediately revealing their newly mined blocks. We propose a statistical test to analyse each miner’s behaviour in five popular cryptocurrencies: Bitcoin, Litecoin, Monacoin, Ethereum and Bitcoin Cash. Our method is based on the realisation that selfish mining behaviour will cause identifiable anomalies in the statistics of miner's successive blocks discovery. Secondly, we apply heuristics-based address clustering to improve the detectability of this kind of behaviour. We find a marked presence of abnormal miners in Monacoin and Bitcoin Cash, and, to a lesser extent, in Ethereum. Finally, we extend our method to detect coordinated selfish mining attacks, finding mining cartels in Monacoin where miners might secretly share information about newly mined blocks in advance. Our analysis contributes to the research on security in cryptocurrency systems by providing the first empirical evidence that the aforementioned strategic mining behaviours do take place in practice.
\end{abstract}
\begin{document}

\flushbottom
\maketitle
%
%
\thispagestyle{empty}

\section*{Introduction}
Blockchains are  decentralised and distributed systems, where sequential, verified data in blocks of a chain and securing data transmission from manipulation through cryptography. Among all the blockchain-based technologies, cryptocurrencies are the most famous ones. The original crypto consensus mechanism is called ``Proof-of-Work''(PoW) and is employed in the majority of cryptocurrency systems\cite{tasca2017taxonomy,10.3389/fbloc.2020.613476}. The consistency of a PoW system's ledger is maintained by all participants solving hash puzzles, a process usually called ``block mining''. In order to solve the puzzles, attempts have to be made through brute force and therefore, \textit{a priori}, the probability of finding a solution is proportional to the number of tries per unit of time a miner is able to perform, measured in hashes per second (H/s). Each miner is then rewarded by a nominal amount of cryptocurrency if they are the first acknowledged miner to find a valid block in the longest chain of the network. This type of rewarding system provides an incentive for miners to contribute their resources to the system, and is essential to the cryptocurrency’s decentralised nature. According to this mechanism, the more mining power (resources) a miner invests, the bigger their chance to mine the next block first\cite{nakamoto2008bitcoin}: as a result miners often join in mining pools to share their mining powers, thus reducing the variance of their rewards. 

PoW mining protocols in principle are tailored to be resistant towards multiple kinds of attacks, but several potentially harmful strategies have been analysed in the literature and some of them have been shown to be profitable under proper conditions. Some attacks might influence the information propagation in the peer-to-peer network, as is the case for Sybil attacks, eclipse attacks \cite{alangot2020decentralized} and routing attacks\cite{apostolaki2017hijacking}; others could threaten data consistency, such as double-spending attacks \cite{karame2012double} or block withholding attacks \cite{bag2016bitcoin}, which are the focus of this paper. According to the PoW protocol, when miners find a block they should submit it to their peer nodes unconditionally. However, in a block withholding attack, miners could decide to not submit the block, or to postpone submitting it. While in the first case, which is also named as sabotage, there is no direct benefit for the attacker but can harm the other miners, the latter one, which is also known as selfish mining (SM), is potentially profitable for the attacker.

\begin{figure*}
\centerline{\includegraphics[width=5in]{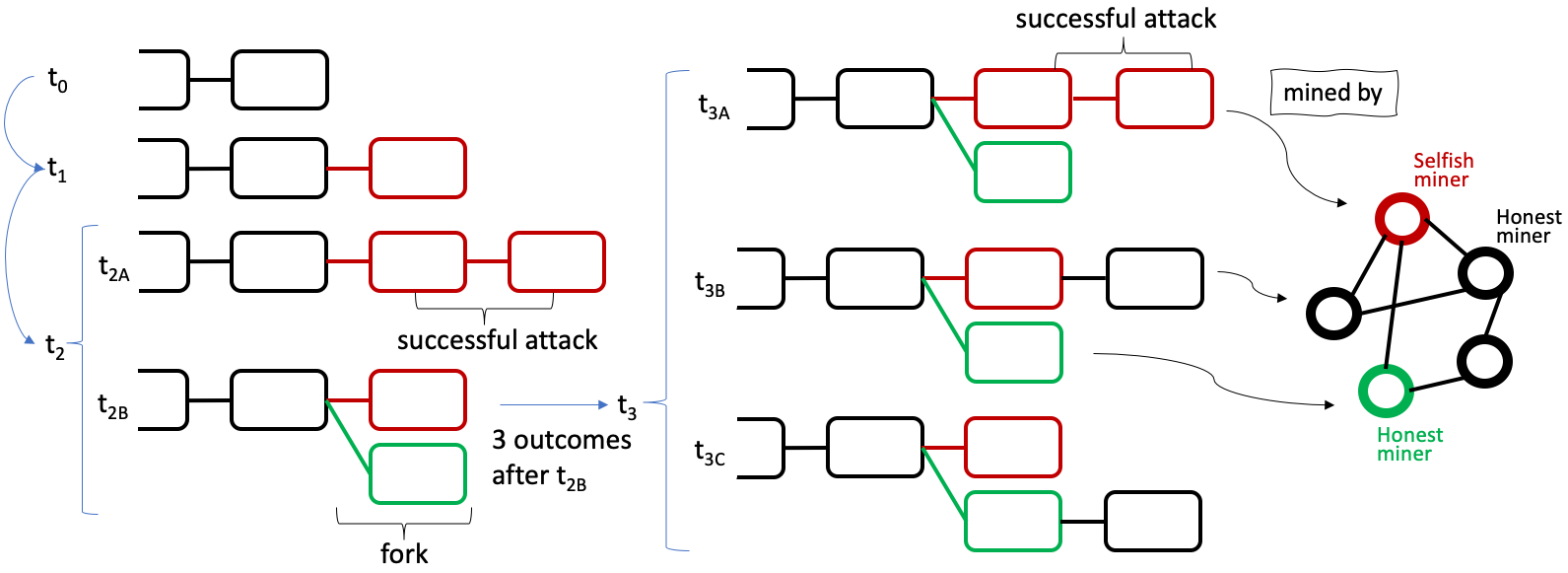}}
\caption{Visualisation of selfish mining strategy.}
\label{Selfish mining introduction}
\end{figure*}

The selfish mining attack was first described by Eyal and Sirer\cite{eyal2014majority} in 2014. They defined the SM strategy as follows: ``the selfish mining pool keeps its mined blocks private, secretly bifurcating the blockchain and creating a private branch. [...] their private branch will not remain ahead of the public branch indefinitely. Consequently, selfish mining judiciously reveals blocks from the private branch to the public, such that the honest miners will switch to the recently revealed blocks, abandoning the shorter public branch.''. The different strategies of SM are shown in Fig.~\ref{Selfish mining introduction}. Let's classify the miners in a stylised P2P mining network in two groups: selfish (red node) and honest (green and black nodes). At $t_1$, a selfish miner mines a block (in red) after the longest chain terminating at $t_0$, but withholds it and secretly mines on the private branch. Then, if the selfish miner finds the next block ($t_{2A}$), the attack is successful and they can choose whether to publish the new chain or continue mining selfishly; however, if a honest miner also finds a block (in green) at the same height ($t_{2B}$), the selfish miner will immediately publish its secret block, and there will be a competition. Later ($t_{3}$, $t_{3A}$), if the selfish miner finds the next block after its own block, it is also a successful attack, whereas if ($t_{3B}$) an honest miner finds the next block after the selfish miner’s block, the selfish miner still enjoys the revenue of the first block. The only negative outcome for the selfish miner is ($t_{3C}$), where an honest miner finds the next block after the honest miner’s block, resulting in the selfish miner gaining nothing. Eyal and Sirer point out another problem caused by the SM attack which is the waste of resources by honest miners on the shorter public branch: if the rewards of selfish miners encourage more honest miners to join the selfish mining pool, it may eventually lead to the selfish pool holding the majority of mining power (51\% attack) and the failure of the cryptocurrency ecosystem. 

The formulation of the SM attack has drawn a lot of attention and many extended mining strategies have been proposed, such as \textit{stubborn mining} \cite{nayak2016stubborn} and the \textit{publish-n strategy}~\cite{liu2018strategy}. Many of these extended strategies use Markov Decision Processes (MDP) solving to compute optimal selfish mining strategies and have managed to lower the profitability threshold of running a SM attack from 25\% hash power ~\cite{eyal2014majority}, to 23.21\%~\cite{sapirshtein2016optimal}, or even 21.48\%~\cite{bai2019deep}. Meanwhile, various countermeasures have been proposed against SM attacks. Zhang \cite{zhang2017publish} has categorised the existing defence methods into two approaches: 1) making fundamental changes to the block validity rules, as for example adopting the ~ \textit{ZeroBlock}~\cite{solat2017brief} timestamp-free solution which requires that each block must be generated and received by the network within a maximum acceptable time interval, or 2) lowering the chance of honest miners working on the selfish miner’s chain during a forked situation, as in the case of the \textit{Freshness Preferred} defence \cite{heilman2014one} which uses unforgeable timestamps issued by a trusted party, providing an incentive for miners to immediately publish newly mined blocks.
To replace the original Bitcoin Fork-Resolving Policy (FRP), denoted by \textit{length FRP}, Zhang \cite{zhang2017publish} proposed \textit{weighted FRP}~: it asks miners to compare the weight of the chains instead of their length, where the \textit{weight} is the number of ``in time" blocks in the chain, and a block is considered ``in time"  based on an upper bound on the block propagation time. However selfish miners' timely reaction to another competitive block, and the high cost of changing the blockchain's fundamental design, might be obstacles to efficiently implement the defence against SM attack. More essentially, the problem of how to detect these selfish miners and quantify the size of the attack is a more urgent problem for the already running blockchain platforms. Recently, in the end of 2020, Nicolas summarised 20 primary selfish mining attack countermeasures using the proposed taxonomy of defensive strategies\cite{nicolas2020blockchain}, and analysed the benefits and limitations of 6 models under his detection category. From his summary one can easily find that most of the existing detection methods have not been tested on real blockchain systems. A framework using deep reinforcement learning to analyse attacks on blockchain incentive mechanisms, called \textit{SquirRL}, has also been proposed by Hou \cite{hou2019squirrl}. When using SquirRL to evaluate both single and multiple agent selfish mining attack in Bitcoin, Monacoin, Vertcoin and Litecoin, Hou only scraped the estimated total hash power hourly from real cryptocurrencies. However, none of these previous studies has detected selfish miners in any real blockchain platforms. The question of whether selfish mining exists in practice or not is largely left unanswered so far.  Stochastic modelling of the mechanism, has shown that attackers can actually also leverage on their location in the P2P network \cite{9657011,jcp2020016}. 
 
Although selfish mining attacks have not been empirically discovered by academic research, Monacoin, a cryptocurrency developed in Japan, reportedly has suffered a selfish mining attack that caused roughly \$90,000 in damages\cite{saad2019countering}. Therefore, the empirical evidence on whether miners do deviate from the mining protocols in practice is important to the security and stability of cryptocurrencies \cite{bonneau2015sok}, and it is thus necessary to direct research towards refining detection methods. 

In our previous work\cite{li2020mining,li2020proof}, we had tried to identify the selfish miners in real cryptocurrency
systems by using the Miner Sequence Bootstrapping model(MSB), the core of which is to shuffle simulations of the sequence of miners’ block discoveries. Based on this insight, in this paper we propose a more interpretable statistical test to evaluate miners' behaviour in five popular PoW-based cryptocurrencies: Bitcoin, Litecoin, Ethereum, Monacoin and Bitcoin Cash. We hypothesise that selfish miners' behaviour of selectively revealing their mined blocks would cause abnormal probabilities of successive block discovery, diverging from normal behaviour of statistical independence of mining outcomes. As we show in the following, under the null hypothesis of ``honest" mining the probability of observing two blocks in a row mined by the same miner is given by the \textit{type II binomial distribution of order 2}\cite{ling1988binomial}, which we use to construct a statistical test to detect mining anomalies. In order to optimise the detection results, we also apply heuristics-based address clustering techniques on all UTXO blockchain datasets. Furthermore, we extended the object of our method from single miner (mining pool) to pairs of miners that may constitute a \textit{mining cartel}, in which miners secretly share information related to blocks discovery before publishing. Our main contributions are listed as follows : 
\begin{enumerate}
    \item To the best of our knowledge, our empirical research on selfish mining attacks and mining cartels in real cryptocurrency systems is presented for the first time. Mining attack detection is important for maintaining blockchain security and could be a fundamental index for cryptocurrencies ranking in the future.
    \item In our mining behaviour detection test, we use a \textit{type II binomial distribution of order 2} to compute each miner’s probability of successive block discovery. This can be widely applied in various competitive consensus protocols, including but not limited to PoW and PoS. 
    \item Our results show that in some cryptocurrencies abnormal miners do secretly collaborate in mining cartels; this could raise concerns about concentration of mining power which has been ignored by most of previous studies.
    \item We highlight the importance of heuristic address clustering for empirical studies in real blockchain systems, especially for user behaviour analysis.
    \item Our empirical analyses also reveal that mathematical or economical models that focus on cost-benefit analysis could fail to detect some behaviours, as participants of cryptocurrencies might have bounded rationality or be risk seeking.
\end{enumerate}

\section*{Results}
\subsection*{Dashboard of Datasets}
\begin{figure*}[t]
\captionsetup{justification=centering}
  \centering
    \includegraphics[width=.8\linewidth]{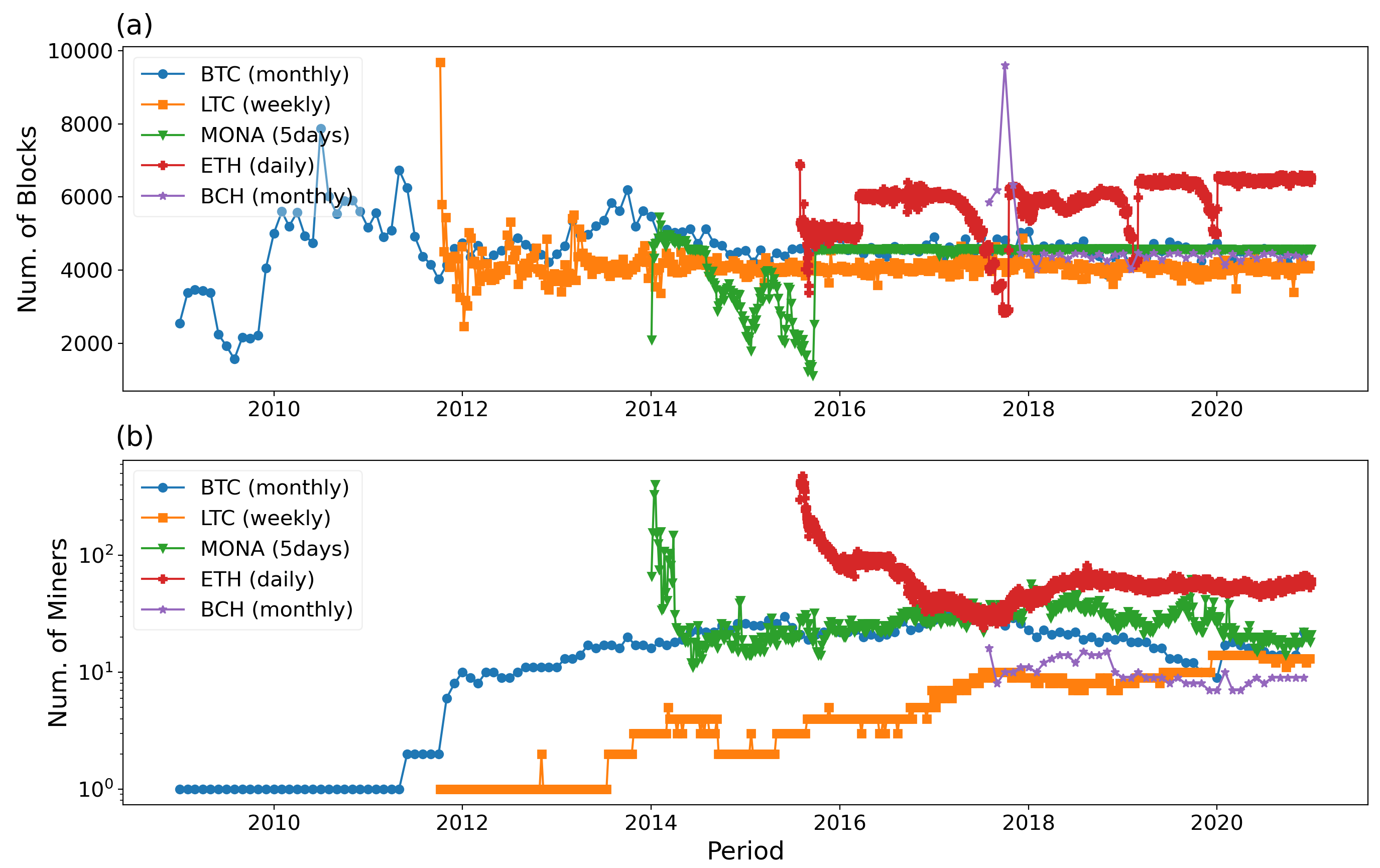}

  \label{fig:num_bm}

\caption{Number of miners and blocks during each period in five cryptocurrencies.}
\label{fig:Num_miner_block}
\end{figure*}

The mining difficulty adjustment in the PoW protocol ensures a fixed average time between each block, called ``block time''. Since Bitcoin (BTC), Litecoin (LTC), Monacoin (MONA), Ethereum (ETH) and Bitcoin Cash (BCH) have different block times, in order to have compatible datasets we split the blockchain in different time intervals tailored to maintain a similar number of blocks ($\sim$ 5000) in each sample. This amounts to monthly (BTC and BCH), weekly (LTC), 5 days (MONA) and daily (ETH) splits. In Fig. \ref{fig:Num_miner_block} we show the number of blocks mined in each time interval for the five cryptocurrencies from the genesis block until the end of our dataset on December 2020. One can find that the mining markets of all the five cryptocurrencies have unstable stages of block mining with different lengths after launch. However, because of the difficulty adjustment, the amount of blocks in each stated period is similar in five coins (as shown in Fig.\ref{fig:Num_miner_block}(a)). In our Ethereum and Monacoin datasets we only have information about each block miner's address, while miners' addresses were tagged to named mining pools in the Bitcoin, Litecoin and Bitcoin Cash datasets. 

\begin{figure*}[t]
\centering
\includegraphics[width=\textwidth]{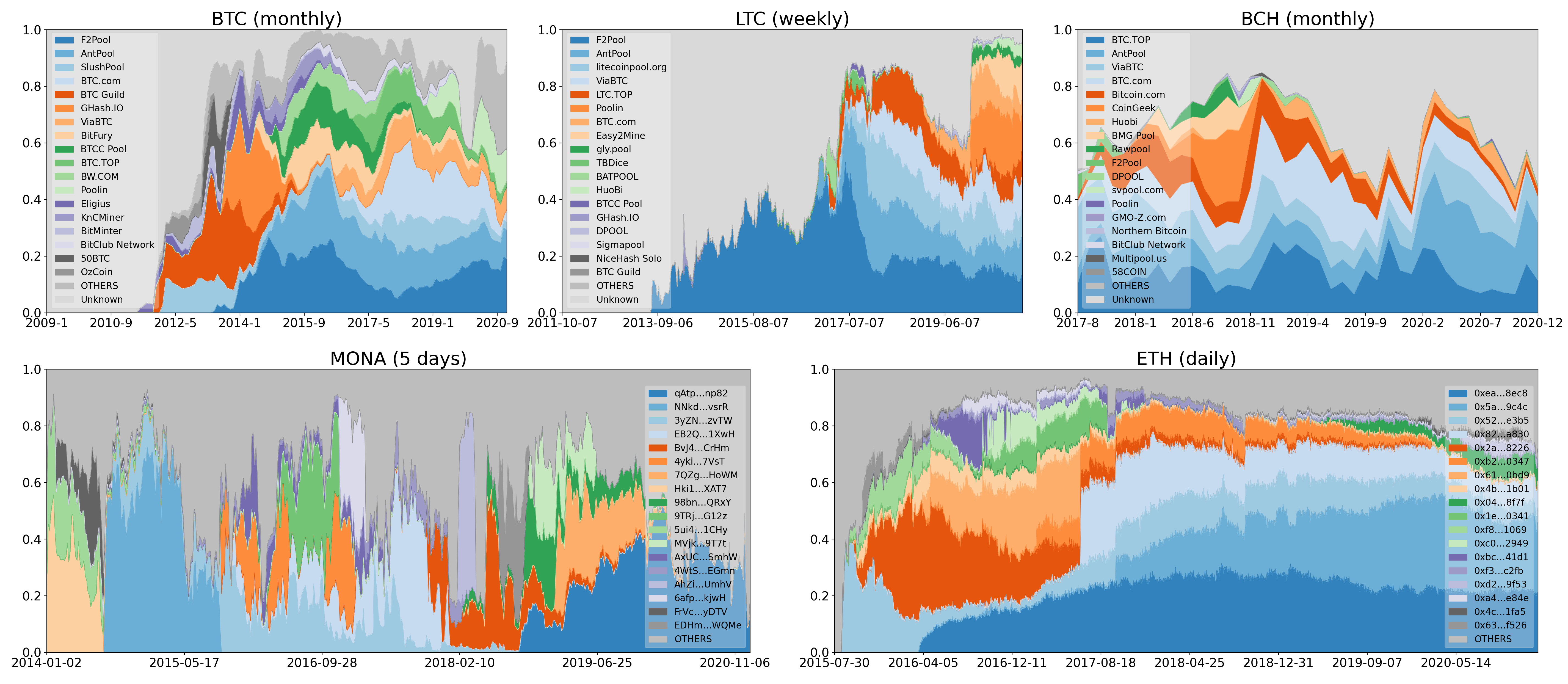}
\caption{Periodic hashing power (share of blocks) distribution in BTC, LTC, BCH, MONA and ETH.}
\label{fig:Power_stacked}
\end{figure*}

We further show the revenue (amount of mined blocks) distributions in each period in Fig. \ref{fig:Power_stacked}. The ``Unknown'' miner are some mining addresses whose identities cannot be traced back to any known pool. It is worth mentioning that some of the unknown mining addresses might be owned by named pools (e.g. to hide their activities such as selfish mining). In detail, one can observe that in BTC and LTC as time flows more and more blocks were mined by named pools, while in BCH there are more than 20\% of blocks that are mined by ``Unknown'' miners all the time. Comparing MONA and ETH where we lack the information about mining pool identity, we find that in MONA the revenue distributions among miners' addresses is more volatile.

In addition, according to the ``PoW'' mechanism, the fair proportion of blocks a miner may discover during a time period (i.e. their blocks share) is equal to their share of mining power. Since we lack better estimators of hash rates, for the rest of this paper we use a miner's share of blocks as a proxy for its mining power. 

\subsection*{Address Cluster}
After finding the ``Unknown'' miners and large fluctuation of hashing power distribution, to enhance the datasets we adopt known methodologies to cluster addresses controlled by the same miner\cite{kalodner2020blocksci, vallarano2020bitcoin, campajola2022evolution, campajola2022microvelocity}. All the heuristic methods we applied exploit inherent properties of UTXO-based transactions, which can include multiple inputs and multiple outputs and generate patterns that allow to cluster addresses together. This was not possible on the account-based blockchain of Ethereum, where only one input and one output can appear in the same transaction. Further details of the three applied methodologies (\textit{$H_1$}, \textit{$H_2$}, \textit{$H_p$}) are provided in the Methods section.

\begin{figure*}[t]
\centering
\includegraphics[width=4in]{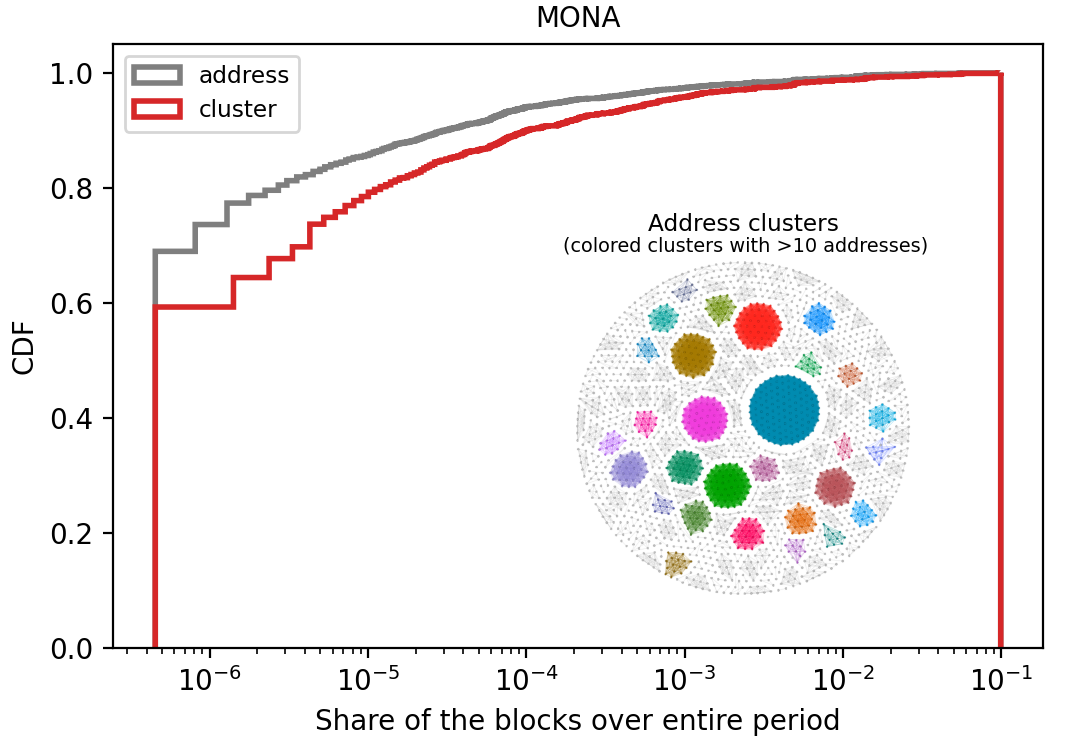}
\caption{Address Clustering in Monacoin}
\label{fig:Monacoin_clus}
\end{figure*}

We apply the most basic method, \textit{$H_1$}, to cluster the miners' addresses in the Monacoin dataset. The distribution of mined blocks share among different entities (addresses or clusters) is shown in Fig. \ref{fig:Monacoin_clus}, and the outcome of clustering in MONA is shown in the subplot of network in Fig.\ref{fig:Monacoin_clus}. In the latter dots are addresses, connected and marked in the same colour if they belong to the same cluster, and the highlighted communities are the larger clusters with more than 10 addresses. It can be seen that the \textit{$H_1$} methodology aggregates miner's addresses into clusters of different size, effectively changing the estimation of hashing power attributed to them.

For the Bitcoin, Litecoin and Bitcoin cash datasets where we already knew the named-pools of part of the blocks, we try all the three mentioned heuristics to tag the ``Unknown'' miners to named pools, with the priority order as $H_1 >H_2 >H_p$. In other words, in each among the BTC, LTC, and BCH datasets, firstly we apply \textit{$H_1$} to cluster the miners' addresses, and then each ``Unknown'' miner whose address could be clustered together with all the addresses of a named mining pool will be tagged to this named pool. We then do the same using the \textit{$H_2$} and \textit{$H_p$} methods to complement the tagging.

\begin{figure*}[t]
\centering
\includegraphics[width=\textwidth]{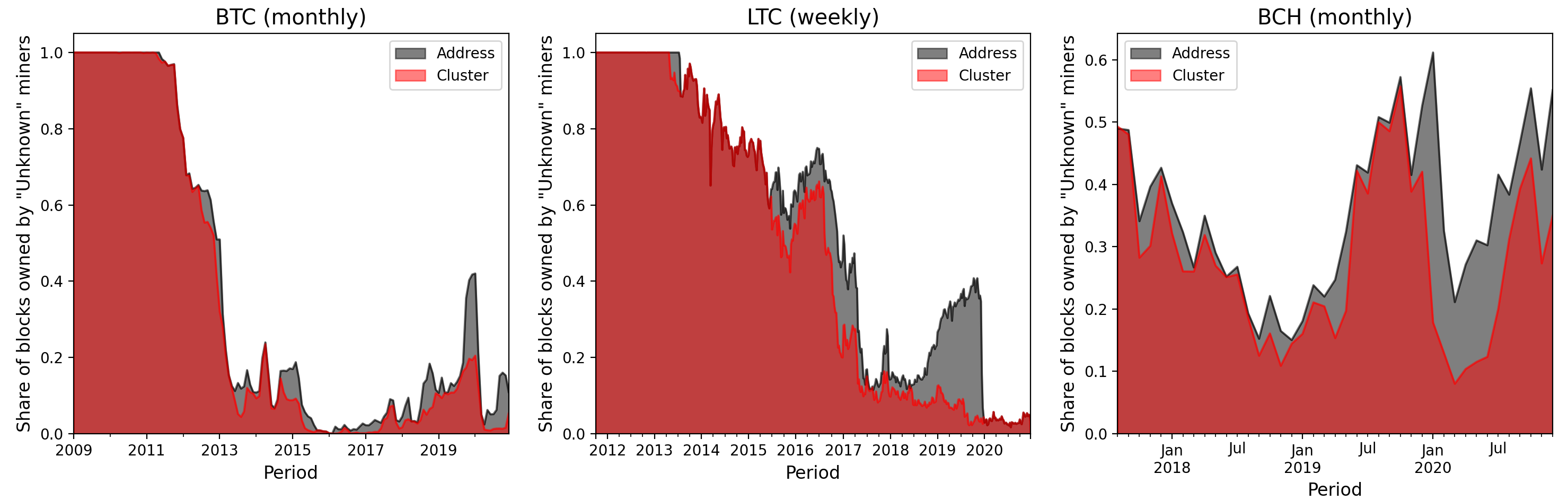}
\caption{Periodic share of blocks owned by ``Unknown'' miners before and after clustering in BTC, LTC, BCH}
\label{fig:Power_stacked_Unknown-clus}
\end{figure*}

The result of this procedure is shown in Fig. \ref{fig:Power_stacked_Unknown-clus}, where it is clear that although it's difficult for address clustering heuristics to tag all the ``Unknown'' miners, a significant fraction of blocks can be attributed to a tagged pool, which is important for a more accurate estimation of miners' actual computing (hashing) power.

\subsection*{Detection of Selfish Miners}
To detect abnormal selfish mining behaviour, we devise a statistical test that we apply on each miner's sequence of mined blocks, the null hypothesis being that miners are ``honest", i.e. they act without selfish behaviour. As we show in the Methods section, under the null hypothesis the event of whether a miner mines a block or not is a Bernoulli random variable, with the success probability equal to the miner's hashing power share. 
However a successful selfish mining attack could lead to anomalies in a miner’s outcome of discovering blocks \textit{in sequence}. Therefore, we design our test statistic to identify suspicious miners by the amount of times in which they mine successive blocks, i.e. the number of success \textit{runs} of length 2, 
whose probability distribution under the null is given by a \textit{type II binomial distribution of order 2} \cite{ling1988binomial}. To account for multiple hypothesis testing errors we apply the Benjamini-Hochberg correction \cite{benjamini1995controlling} for the $p$-values to control for excess false positives, setting the target False Discovery Rate (FDR) to 5\%. 

\begin{figure*}[t]
\begin{subfigure}{.5\textwidth}
  \centering
    \includegraphics[width=3.4in]{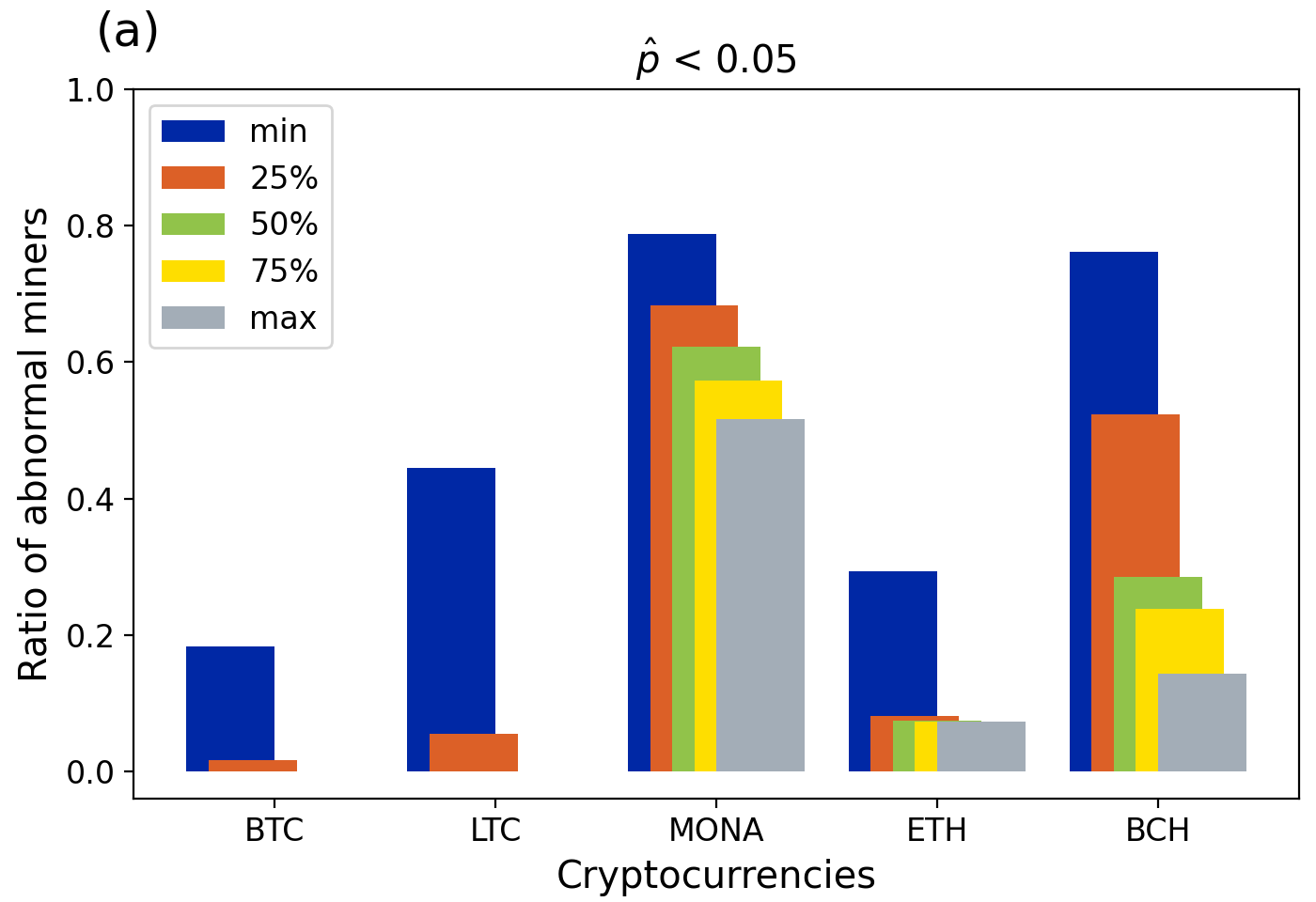}
  \label{fig:SM_5coin}
\end{subfigure}
\begin{subfigure}{.5\textwidth}
  \centering
  \includegraphics[width=3.4in]{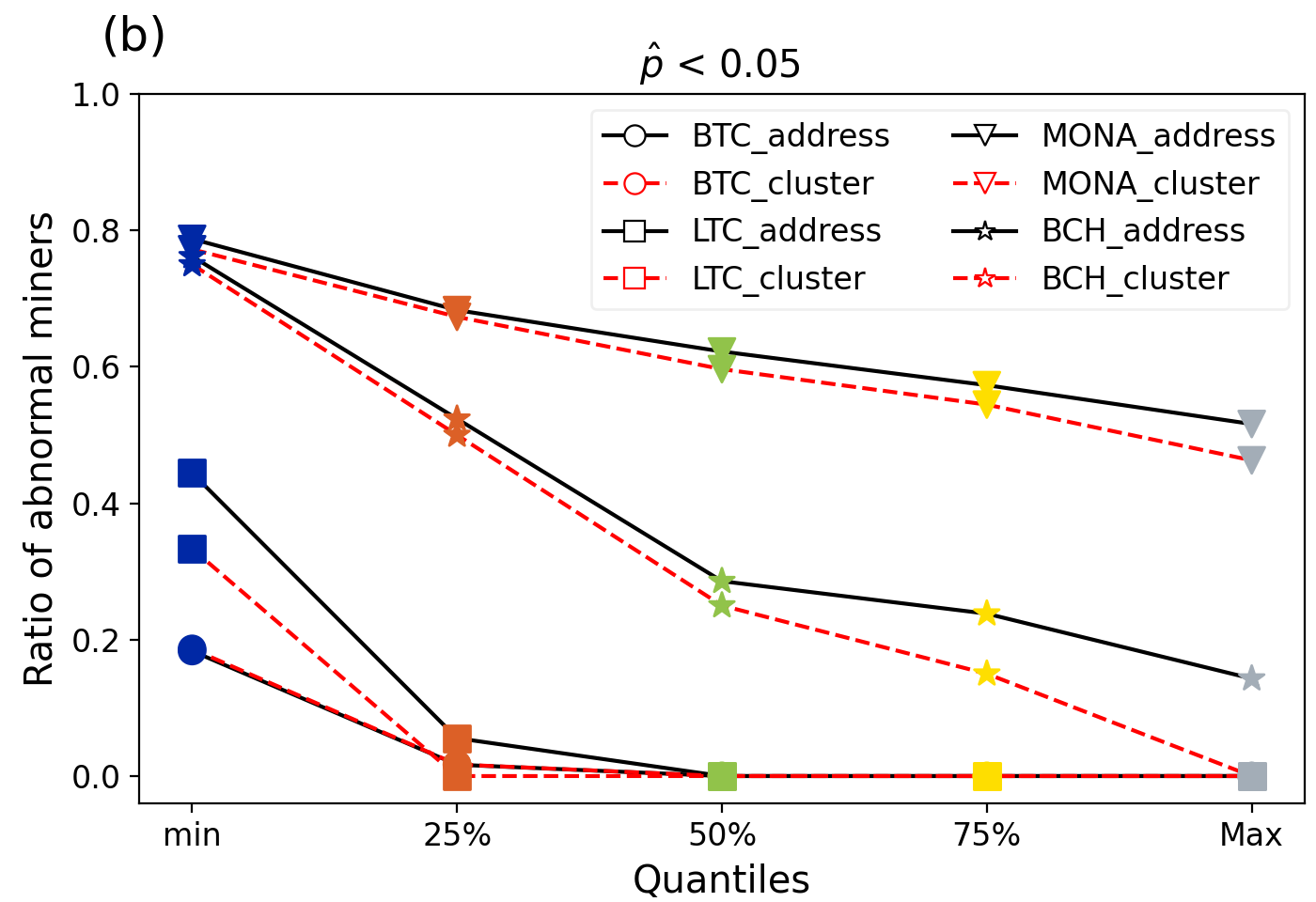}
  \label{fig:SM_clus}
\end{subfigure}
\caption{Ratio of abnormal miners in BTC, LTC, MONA, ETH and BCH. In (a), the bar in different colors respectively shows the percentage of unique miners, whose minimum value, first, second, and third quartile, as well as maximum values of the $\hat{p}$ in different period is less than 0.05. In (b), the red dashed lines display the comparison between results after address clustering and the original results shown in (a) of BTC, LTC, MONA and BCH.}
\label{fig:SM_index}
\end{figure*}

The results of our tests (before address clustering) are shown in aggregate in Fig. \ref{fig:SM_index}. In Fig.\ref{fig:SM_index}a, each bar shows the proportion of abnormal miners (with the corrected $p$-values, $\hat{p} < 0.05$) in the five cryptocurrencies. Bars in different colours represent results under different classification criteria: the blue, orange, green, yellow and the grey bar respectively show the fraction of miners for whom at least 25\%, 50\%, 75\% or all (max) tests on the considered time periods reject the null hypothesis at 5\% FDR. For example, in Monacoin, the result expressed by the grey bar shows that about half of miners have behaved selfishly in all the periods they were active. 

We then compare the detection results before and after address clustering, shown in Fig. \ref{fig:SM_index}b. Following address clustering the ratio of abnormal miners in each coin decreases; however, even after clustering, there were more than 46\% miners who always engaged in strategic mining behaviour in Monacoin. The reduction in abnormal ratio when changing the criterion from lower quartile (25\%) to maximum (max) is larger in Bitcoin cash than in Monacoin, which shows the abnormal miners in Monacoin might be more likely to continuously behave with the selfish strategy, or alternatively that many malicious miners entered Monacoin only to run a SM attack, leaving the system right after.

\begin{figure*}[t]
  \centering
  \includegraphics[width=.8\textwidth]{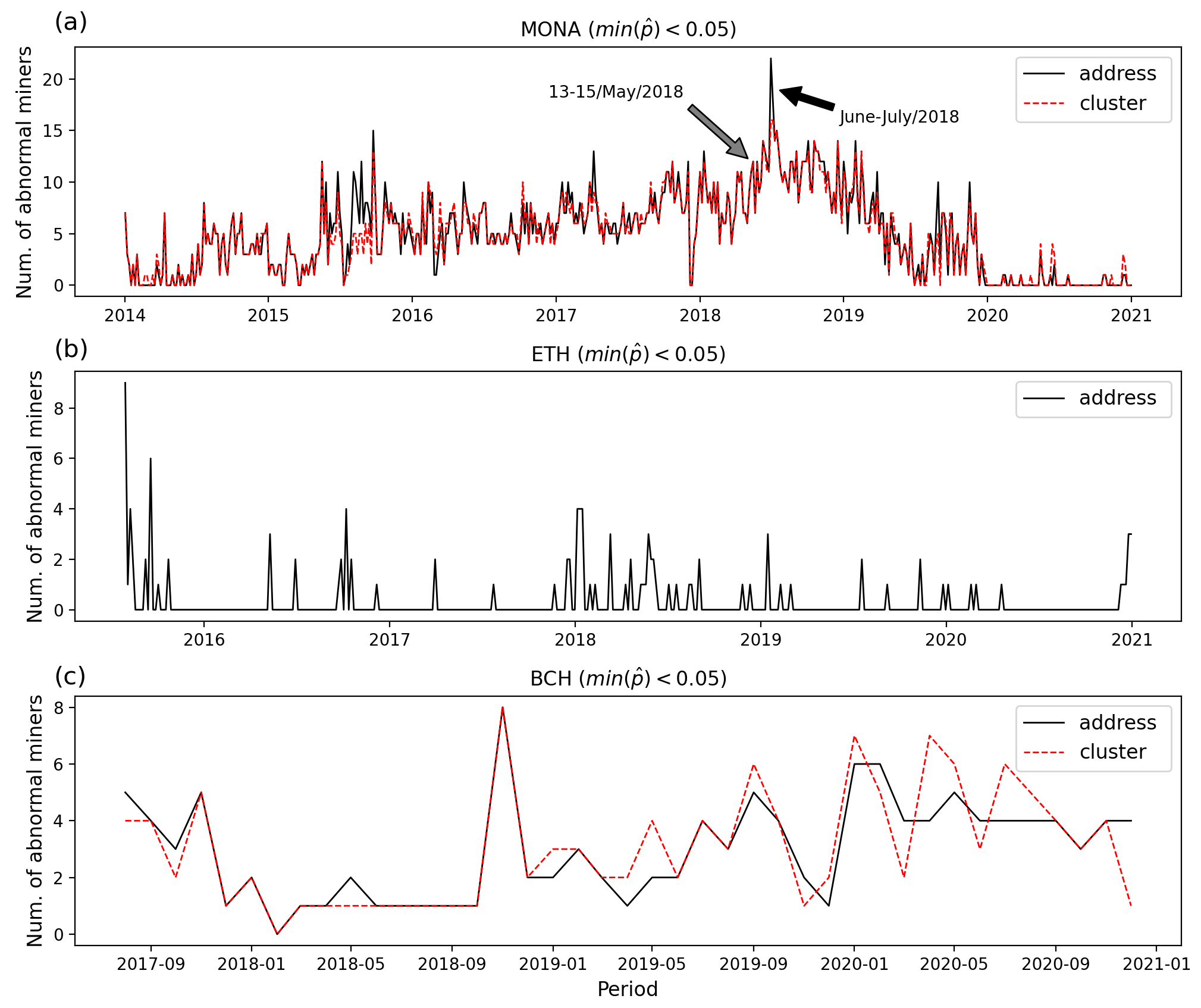}
\caption{Number of abnormal miners during each time period in MONA, ETH and BCH. The black lines present the results before doing address clustering, while the red dashed lines show the results after address clustering.   }
\label{fig:SM_period}
\end{figure*}

In addition, we show the number of abnormal miners for each period in Monacoin, Ethereum and Bitcoin cash in Fig. \ref{fig:SM_period}. The result of each period includes all miners whose corrected $p$-value, $\hat{p}$, is smaller than 0.05 in that period. The empirical results of Monacoin (\ref{fig:SM_period}a) show that the period with the most abnormal miners is around June-July 2018, which is near but much longer than the period \textit{13-15 May 2018} when Monacoin announced they had suffered from a selfish mining attack. Besides, a part of miners might have been trying the selfish mining attack throughout time, not only during the mentioned periods. It seems that the selfish mining attack on Monacoin was contained after 2019 as we see a downward trend in the amount of abnormal miners. The result in Fig. \ref{fig:SM_period}c shows that several miners in Bitcoin cash might try to conduct the selfish mining attack much more erratically, and a large number of abnormal miners appeared in Nov. 2018 in Bitcoin cash, with still a few abnormal miners persisting into more recent years. Similarly in Ethereum (Fig.\ref{fig:SM_period}b), there were more abnormal miners at ETH's launch, with SM attacks being more frequent in 2018 and occasionally occurring during the run time. 

\begin{figure*}[t]
  \centering
    \includegraphics[width=\textwidth]{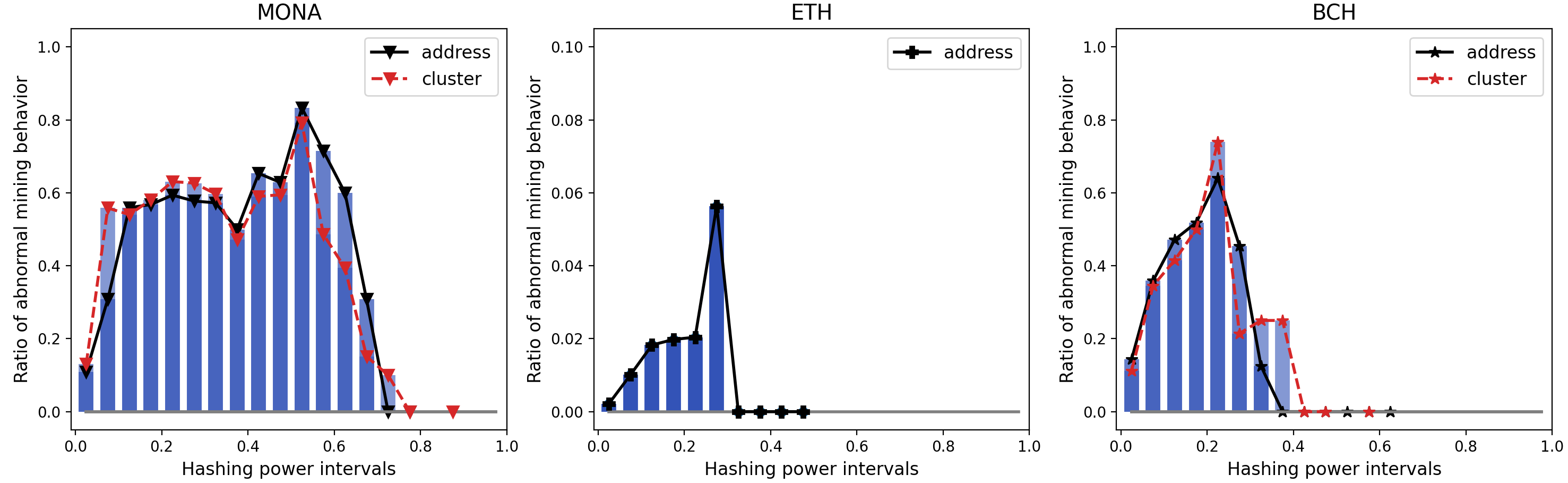}
\caption{Fraction of abnormally behaving miners sorted by hashing power ranges in MONA, ETH and BCH.}
\label{fig:SM_power}
\end{figure*}

To further research the effect of increasing mining power on the potential of doing SM attack, we group active miners in each period by their corresponding hashing power in that period, and calculate the proportion of abnormal miners in each hashing power interval. In Fig. \ref{fig:SM_power}, one can find that in Monacoin the incidence of SM behaviour increases with miners' power when below 50\% hashing power, and this increasing incidence also exists in Ethereum when below ~30\% hashing power, as well as in Bitcoin Cash below ~25\% power.

\subsection*{Network of Mining Cartels}
\begin{figure*}[t]
\begin{subfigure}{.3\textwidth}
  \centering
    \includegraphics[width=1.5in]{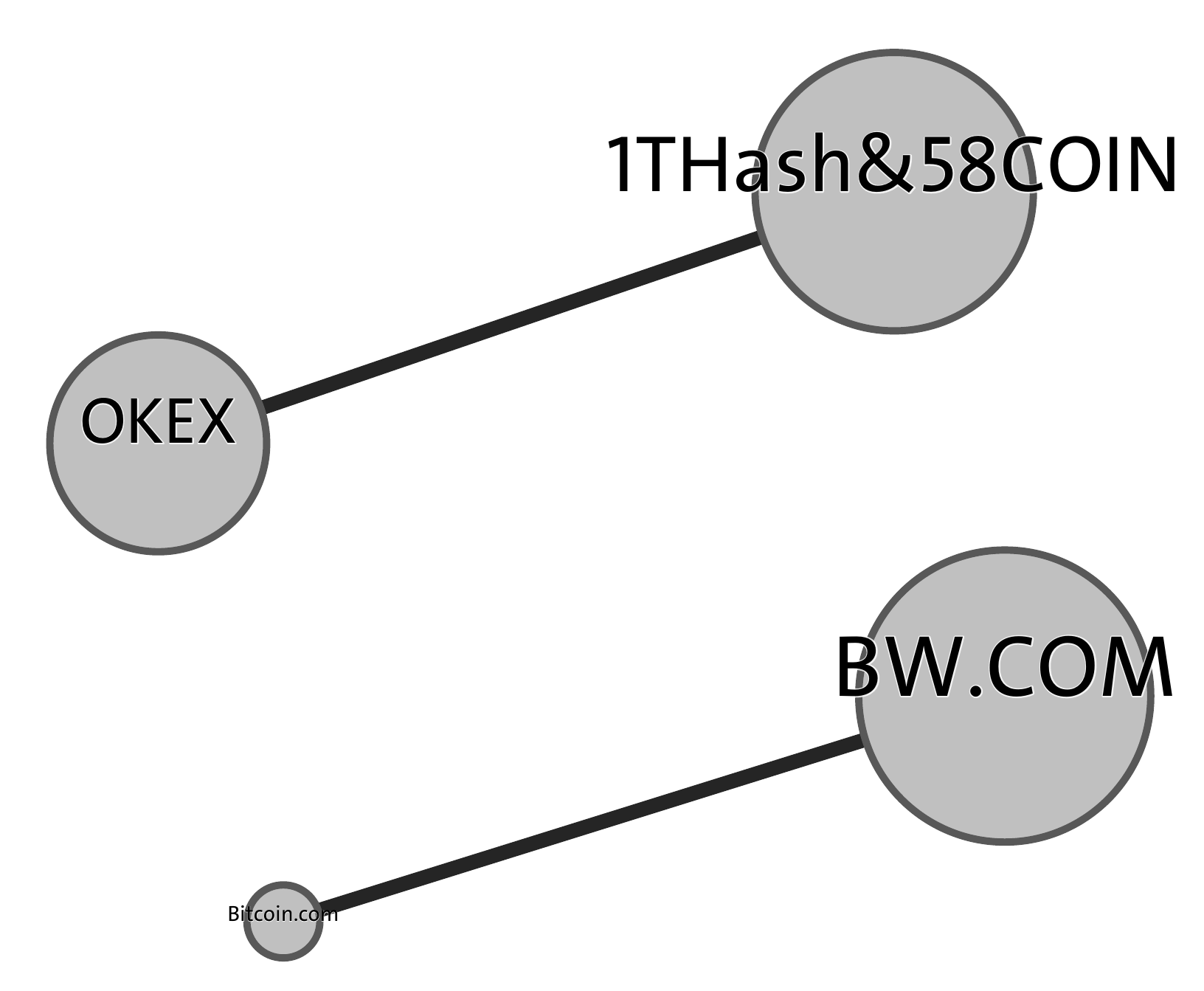}
  \caption{BTC}
  \label{fig:BCT_cartel}
\end{subfigure}
\begin{subfigure}{.3\textwidth}
  \centering
  \includegraphics[width=1.5in]{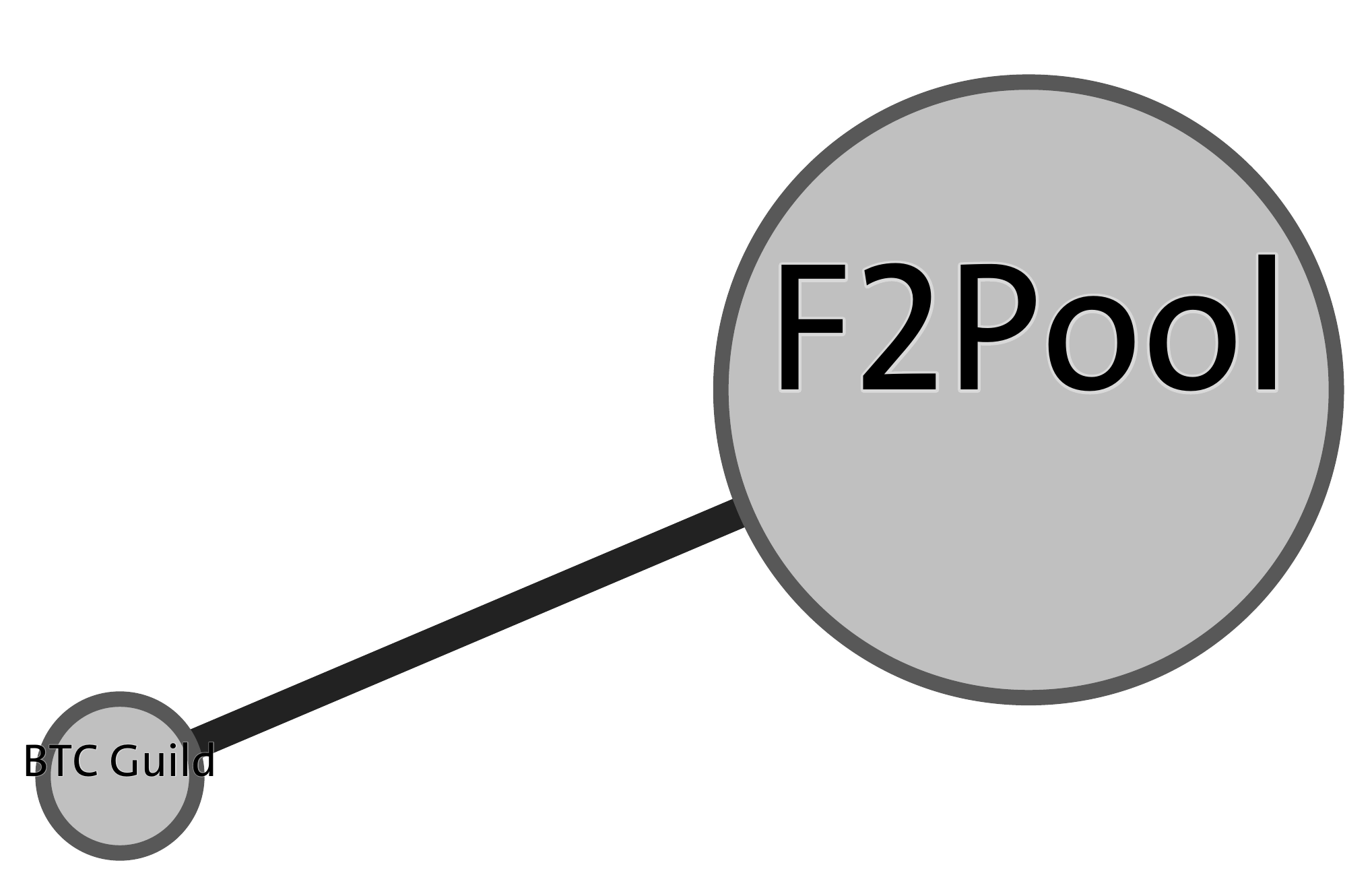}
  \caption{LTC}
  \label{fig:LTC_cartel}
\end{subfigure}
\begin{subfigure}{.4\textwidth}
  \centering
  \includegraphics[width=2in]{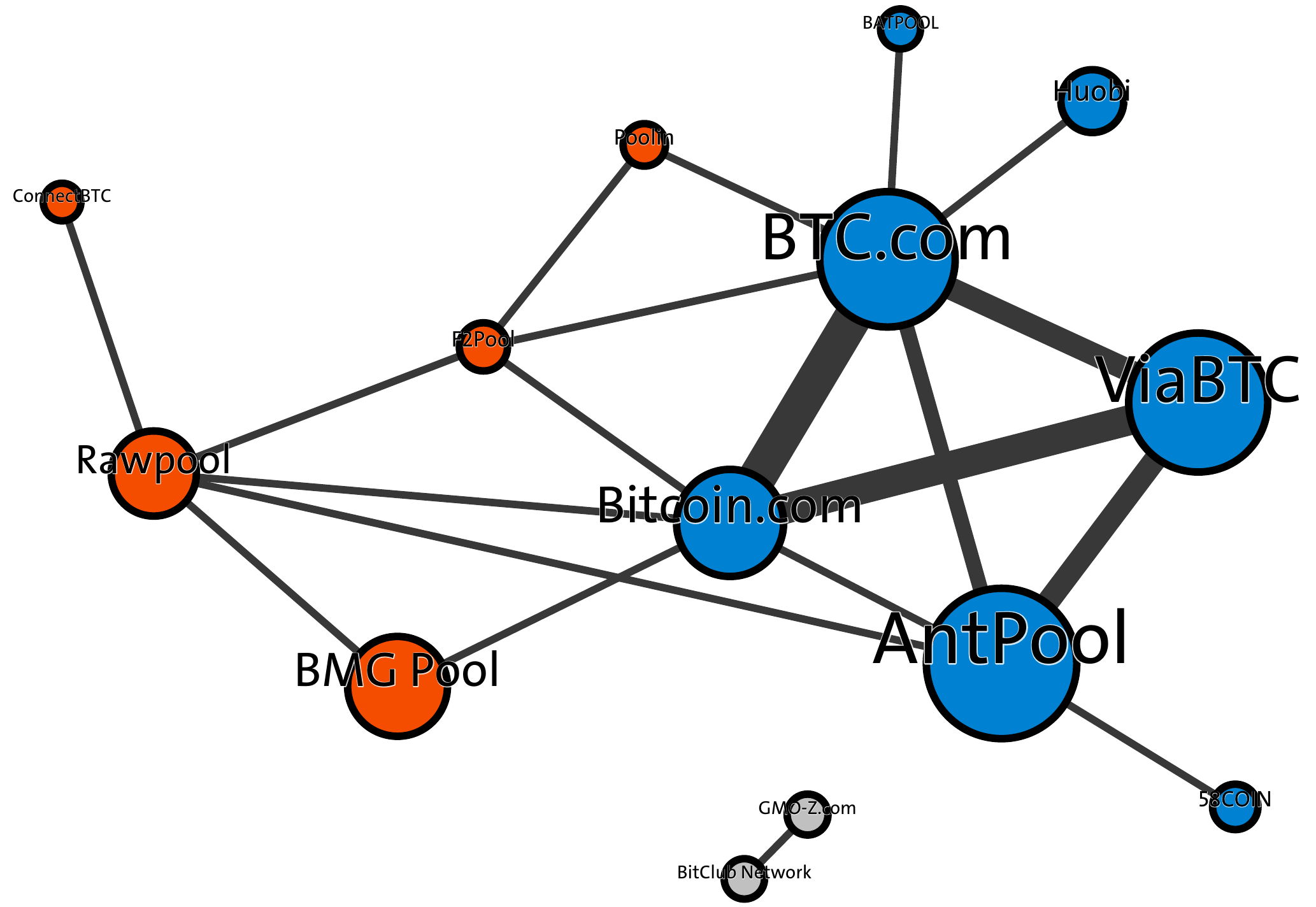}
  \caption{BCH}
  \label{fig:BCH_cartel}
\end{subfigure}

\begin{subfigure}{.5\textwidth}
  \centering
  \includegraphics[width=3.2in]{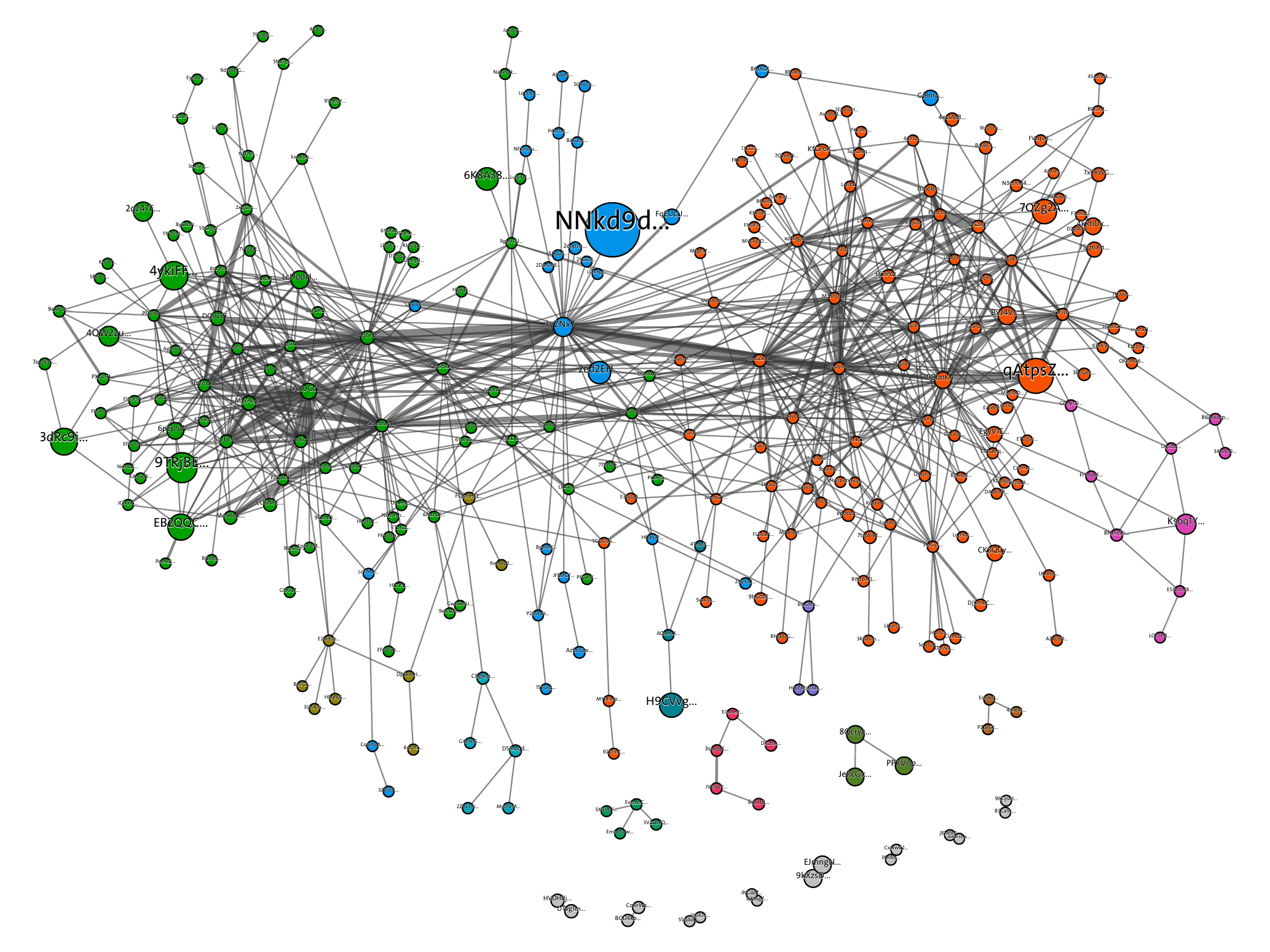}
  \caption{MONA}
  \label{fig:MONA_cartel}
\end{subfigure}
\begin{subfigure}{.5\textwidth}
  \centering
  \includegraphics[width=3.2in]{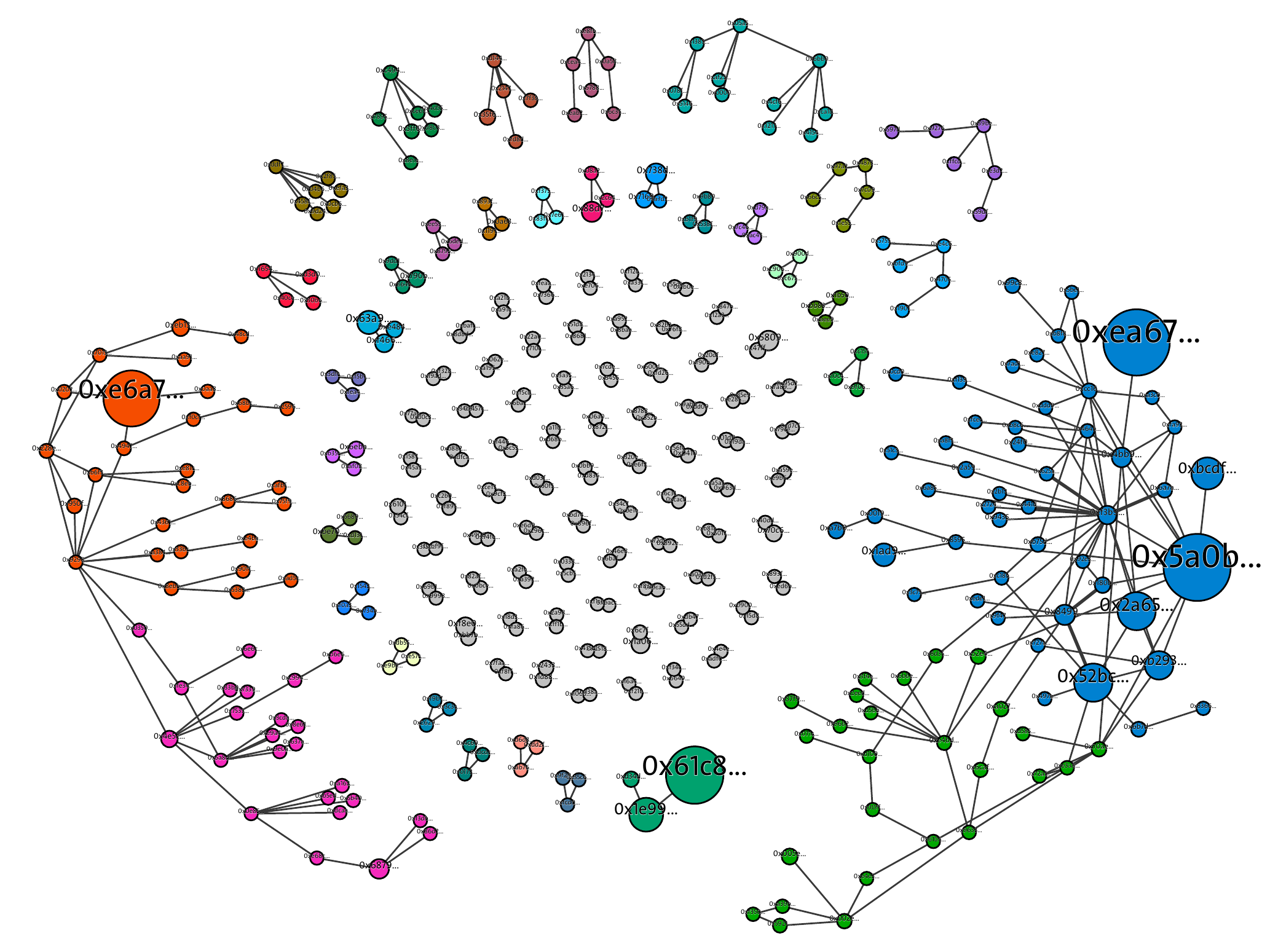}
  \caption{ETH}
  \label{fig:ETH_cartel}
\end{subfigure}

\caption{Networks of Mining Cartel in BTC, LTC, BCH, MONA and ETH. Each node represents a pool (in BTC, LTC and BCH) or an address (in MONA and ETH), and each link represents an identified cartel between two miners. The weight of each link is the number of times the pair of miners has been detected as a cartel in all periods. }
\label{fig:Cartel_5}
\end{figure*}

In order to detect the existence of a mining cartel where different miners share the information in advance among themselves and perform a coordinated selfish mining attack, we have extended our methods from testing single miners to pairs of miners. Considering pairs of miners $i$ and $j$ as a group $ij$, we conduct the similar hypothesis tests as above for each pair of miners and also calculate their corrected $p$-values, $\hat{p}_{ij}$ in each period. Then we consider the pairs with $\hat{p_{ij}}<0.05$ (but both $\hat{p}_{i}$ and $\hat{p}_{j}$ are greater than 0.05 in the given period) as potential cartels composed by miners $i$ and $j$. After testing each pair of miners in five cryptocurrencies, we show the network of identified mining cartel for each cryptocurrency in Fig. \ref{fig:Cartel_5}, where in each network a link represents an identified cartel between two miners and the weight (width) of the link is the number of times this pair of miners were detected as a cartel in different periods. The size of the node reflects the miner's average hashing power among all his active periods.

As shown in Fig.\ref{fig:BCT_cartel} and \ref{fig:LTC_cartel}, we find two abnormal cartels in Bitcoin and only one in Litecoin, and each of these cartels only includes two members. In Bitcoin cash as shown in Fig.\ref{fig:BCH_cartel}, there are a few cartels, most of which are in a connected subgraph, and the four most powerful mining pools AntPool, BTC.com, ViaBTC and Bitcoin.com (the four biggest blue nodes) are fully connected with each other. We identified many abnormal cartels in both Monacoin (in Fig. \ref{fig:MONA_cartel}) and Ethereum (in Fig. \ref{fig:ETH_cartel}), but the connectivity of cartel networks in Monacoin and Ethereum is totally different. In Monacoin, a bit like in Bitcoin cash, we find large cartels containing two or three powerful miners and many small miners, with a very high connectivity. In addition, there are also some separated small cartels with a few miners. However, the whole cartel network of Ethereum has a low connectivity. There are two separated large cartels, each of which contains several powerful miners, as well as a few cartels of varying size and with a generally low connectivity. 


\section*{Discussion}
The ledger of cryptocurrencies is always maintained through distributed consensus. Proof-of-work (PoW) is the most widely used consensus mechanism, maintaining the consistency of the system’s ledger by requiring validators to solve an arbitrary mathematical puzzle to earn the right to verify transactions. Following the tremendous increase in market capitalisation of cryptocurrencies these years, defence from potential attacks on blockchain system has become an important topic. We considered the problem of selfish mining, one of the attacks which breaks information symmetry in blockchain systems, proposed by Eyal and Sirer in 2014. When employing the selfish mining (SM) strategy, malicious miners selectively keep their newly mined blocks temporarily private instead of publishing them immediately. To our knowledge, most of the previous studies on detection of selfish mining attacks are analytical models without empirical tests on real blockchain systems. 

In this study, we proposed a statistical method to conduct empirical research on detection of selfish miners in five ``PoW''-based cryptocurrencies, namely Bitcoin (BTC), Litecoin (LTC), Monacoin (MONA), Ethereum (ETH) and Bitcoin cash (BCH). Regardless of whether SM actually leads to monetary gains or not, we emphasise that the strategy could lead to anomalies in the frequency with which a miner discovers successive blocks. We also investigated mining cartels, where miners secretly share information about new blocks among partners to pursue a collective SM strategy. Given the fact that existence of mining cartels may cause certain threats to the security of blockchain-based systems but has been ignored in many previous studies, we also proposed our methods to test miners pairwise, in order to detect at least some potential cartels. Our results suggest that although the SM strategy was proposed as an attack to the Bitcoin system, it was employed by more miners in Monacoin and Bitcoin cash. In particular in Monacoin there are about 50\% potential selfish miners. Our detection results are consistent with Monacoin’s own report about having suffered selfish mining attacks. We also detect more mining cartels in Monacoin, Ethereum and Bitcoin cash compared with the two in Bitcoin and only one in Litecoin. The cartel network in Monacoin has a very high connectivity, but the cartels in Ethereum are more separated and most of them are in a tree structure. In addition to that, our results also show the importance of address clustering when conducting empirical studies in real blockchain systems.

There are some limitations to our work. First of all, selfish mining attacks and forming mining cartels are only two of the possible reasons for the anomalies we detect in miner’s rates of successive block discoveries; alternatives include for example finite diffusion times\cite{decker2013information}. Secondly, 
we relied on the empirical frequency of mined blocks to estimate miners' hash rates: while this is the best estimator we can obtain from blockchain data, it is not necessarily accurate.

Our next step is to analyse the miner's (validator) selfish behaviour in cryptocurrencies that apply other consensus mechanisms \cite{neuder2019selfish}. In addition, a further empirical research based on this paper is whether an ``uncle block reward" could cause Ethereum to be more vulnerable to selfish mining attacks \cite{feng2019selfish,ritz2018impact}. Finally, our methods can also be applied as a forensics tool to characterise strategic mining behaviours, contributing to monitoring the security in current cryptocurrency ecosystems.

\section*{Methods} \label{sec:methods}

\subsection*{Anomalies in Selfish Mining Attack}
Following the ``PoW" protocol, a miner's discovery of each block should be random and independent without any influence from the previous blocks, if the information diffuses through the network instantaneously \cite{decker2013information}. Thus, during a certain time period where each miner's hashing power $h_{i}$ is assumed constant, the event whether miner $i$ mined block $t$ or not follows a Bernoulli distribution with probability $h_{i}$. However, when doing a strategic mining attack (e.g. selfish mining), the miners selectively publish their mined blocks to keep their leading height in block competition. This could lead to identifiable anomalies in statistics of successive blocks discovery.

How many consecutive blocks an attacker could mine is important to blockchain security and also to attack detection. As we can imagine when selfish miners keep mining on their private chain, they also take risks of losing the expected revenue. In the competition of solving hash-puzzle, in order to ensure a fair revenue, most of the attacks in PoW systems won't have a long private chain. According to the strategies of selfish mining \cite{eyal2014majority}, when the private chain falls behind the public chain or the lead drops to 1, the attacker will immediately publish their private block. Furthermore, in the research on the alternative ``stubborn mining"\cite{nayak2016stubborn}, the authors did not observe any case where a selfish miner could earn more revenue if they don't merge with public when they fall behind by more than 1 block. In addition, there is a heuristic using the NS3 bitcoin simulator to detect malicious miners by observing the fork height \cite{chicarino2020detection}. The results show that if the mean height of the fork is higher than 2, the blockchain system can be considered under selfish mining attack. All the arguments above then indicate that the length of a private chain would not be very long, usually no more than 2 blocks. 

Therefore, in this paper the statistical analysis of selfish mining behaviour focuses on the case of the same miner mining \textit{two} consecutive blocks. Although a selfish mining strategy may not significantly increase the proportion of blocks mined by strategic miners ~\cite{wright2017fallacy}, using our methodology we can detect the abnormal miners by testing the probability of miners' successive block discovery.

\subsection*{Probability of Successive Block Discovery} 
Assume there are $N$ miners in the system, identified by index $i = 1,\dots,N$. Define a random variable $X(t)$, where $t=1, \dots, T$ is the block index and $X(t)=i$ if miner $i$ mined block $t$. Assuming over the given time period the miner's hashing power $h_{i}$ is constant, $X(t)$ is characterised by a multinomial probability distribution with unit size, $X(t) \sim \mathcal{M}(h,1)$ where $P[X(t) = i] = h_i$ is proportional to miner $i$'s hashing power and, of course, $\sum_i h_i =1$. The auxiliary random variable $Y_i(t)$ which is $1$ if $X(t) = i$ and $0$ otherwise then follows a Bernoulli distribution with probability $h_i$. We can then define our test statistic as the number of times $c_i$ that the event $Y_i(t) = Y_i(t+1) = 1$ has occurred, i.e. that miner $i$ has mined $c_i$ consecutive pairs of blocks among the $T$ total blocks. 

The probability distribution that characterises $c_i$ is given by Ling \cite{ling1988binomial}. Indeed the random variable $c_i$ follows what is called a {\it type II binomial distribution of order 2}, and the expression below (Equation ~\ref{BD_Probability} ) for the probability mass function is also given in the original paper. Calling $c_i^{(T)}$ the random variable $c_i$ for a sequence of length $T$,

\begin{equation}
P(c_i^{(T)} = x) = \begin{cases}
h_i^T & \text{if} \; x = T - 1 \\
2h_i^{T-1}(1-h_i) & \text{if} \; x = T - 2 (>0)\\
\sum_{j=1}^{x+2} h_i^{j-1} (1 - h_i) P(c_i^{(T-j)} = x - \max \lbrace 0, j-2 \rbrace) & \text{if} \; 0 \leq x < T-2
\end{cases}
\label{BD_Probability}
\end{equation}

\begin{figure*}
\begin{subfigure}{.5\textwidth}
  \centering
    \includegraphics[width=3.1in]{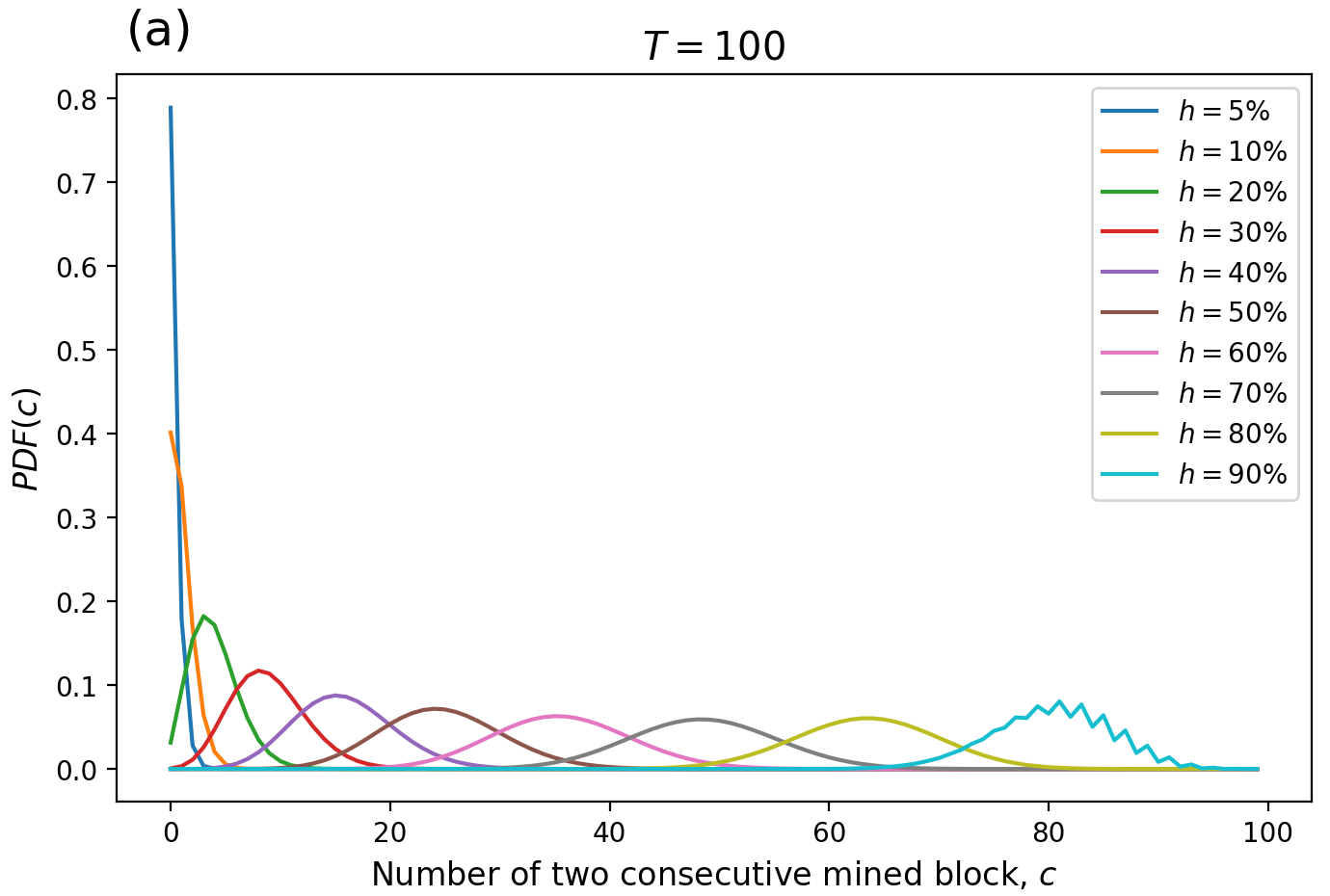}
  \label{fig:Density}
\end{subfigure}
\begin{subfigure}{.5\textwidth}
  \centering
  \includegraphics[width=3.1in]{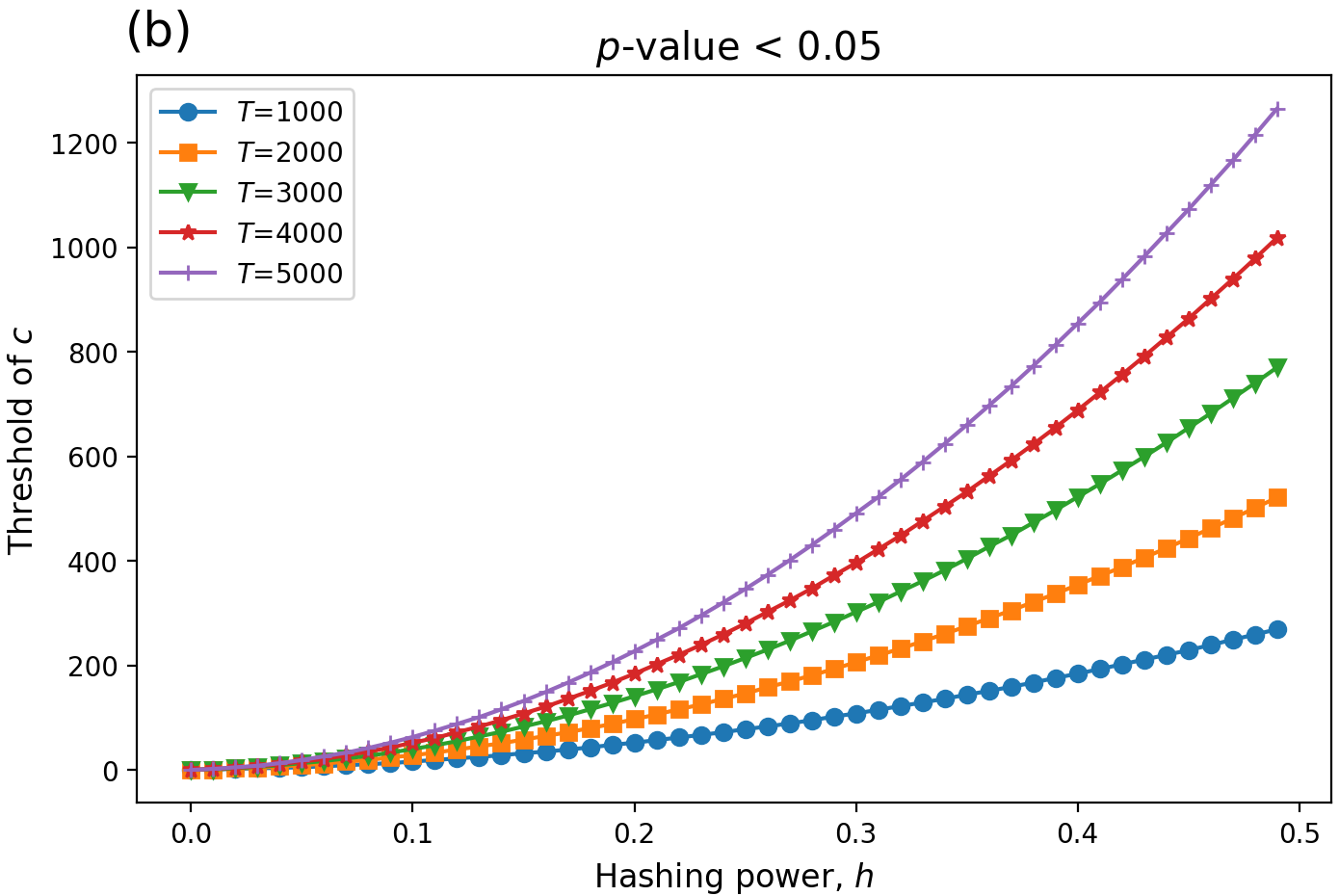}
  \label{fig:Threshold}
\end{subfigure}
\caption{Illustrative diagrams of the statistical method used to detect abnormal mining behaviour in Proof-of-Work protocols}
\label{fig:Statistical}
\end{figure*}

In applying the above formula, Ling also put $P(c_i^{(T)} = 0)=1$ if $T<2$. In Fig. \ref{fig:Statistical}a, we show an example that probability distribution of variable $c$ under different hashing powers $h$ where the amount of blocks $T$ is 100. The most probable value of miner's runs $c$ increases with their hashing power, where a run is defined as two consecutively mined blocks.  

\subsection*{Detection of Abnormal Miners}
One of the main purposes of this paper is to obtain empirical evidence about whether selfish mining behaviours occur in practice or not. To achieve this purpose, we conducted hypothesis tests for every miner in various cryptocurrencies under the null hypothesis that the miner is honest, such that rejections of the null would identify a potential selfish miner in a certain period. Our null hypothesis means that miner $i$ acts non-selfishly in compliance with the protocol, i.e. all blocks are mined randomly and independently. 
Under this null, the $p$-value is going to be the probability that miner $i$ has at least $c_i$ runs of two consecutively mined blocks occurring in a sample of $T$ blocks. Thus, the $p$-value corresponding to the observation of $c_i$ consecutively mined block pairs is $p_i= P(x \geq c_i)$, or
\begin{equation}
p_i =\sum_{x = c_i}^{T-1} P(c_i^{(T)} = x)=1- \sum_{x =0}^{ c_i-1} P(c_i^{(T)} = x)
\label{BD_Pvalue}
\end{equation}

To give better intuition, in Fig. \ref{fig:Statistical}b, we report the critical values $c^*$ of the number of consecutively mined blocks at significance level $\alpha= 0.05$, for different values of mining power $h$ and amount of blocks $T$. That is to say, for example in the purple line, when the amount of the block is 5000 , the miner with less than 30\% hash power but more than 491 runs of two consecutive blocks might conduct strategic mining behaviours whitin the 95\% confidence interval. 


When running multiple hypothesis tests the probability of obtaining one or more false positives (in this case identifying honest miners as abnormal) quickly becomes very high. For this reason it is important to adjust the $p$-values of each test to control for the False Discovery Rate (FDR), i.e. the expected fraction of false rejections among all rejected null hypotheses. We then adjust the $p$-values according to the procedure by Benjamini and Hochberg (BH) \cite{benjamini1995controlling}, where 
the corrected p-value reads
\begin{equation}
\hat{p}_k =\frac{p_{k} * T}{k}
\label{BD_FDR}
\end{equation}
with $p_k$ being the $k$-th smallest $p$-value out of $T$ total $p$-values in the test. 
After getting the $\hat{p}$ for all the miners, we can reject the null with a $5\%$ FDR for miners $k < k^*$, where $k^*$ is the maximum $k$ such that $\hat{p}_{k^*} < 0.05$, i.e. our results is expected to return a $5\%$ rate of false positives out of all rejected nulls.

\subsection*{Detect Mining Cartels}
Either an individual miner or a mining pool doing strategic mining could be considered as a single attacker. However, if some attackers share information earlier or only among themselves and collaborate to achieve the attack, they will be seen as a cartel. A mining cartel is then a secretly coordinating group where miners get together and share timely information related to the blocks mined. In the previous papers \cite{li2020mining,li2020proof}, Li et al. already pointed out that the existence of mining cartels has always been ignored so far. Considering a mining protocol secure as long as the pool's mining power is limited below a certain threshold always relies on the assumption that the miners (or pools) are operating independently. However, strategic miners may have incentives to associate in cartels, such as to benefit from the increased mining power and having information in advance about the blocks mined by the other members. On the other hand, detection of mining cartels contributes to revealing the potential relationships between attackers.  

Based on the assumption that the collaboration in a cartel will cause the same anomalies of successive blocks discovery by cartel members, we run our test method on pairs of miners to detect potential cartels in five cryptocurrency systems. Therefore, we would like to verify whether a cartel has formed between two miners, $i$ and $j$, by measuring the anomalies in their consecutive blocks’ statistics. Specifically, we use $c_{ij}$ which is the number of times that two consecutive blocks is mined by the pair of miners $i$ and $j$ (regardless of the order), to replace $c_{i}$ in Eq. \ref{BD_Probability}. Likewise, we replace $h_{i}$ by $h_{ij}=h_{i}+h_{j}$ which is the estimated aggregated mining power. As a result we can calculate the $p$-value of a pair of miners $i$ and $j$, $p_{ij}$ to which we also apply the usual FDR correction, and then use the corrected $\hat{p}_{ij}$-value to detect the cartel between miner $i$ and $j$. 


Of course one may identify a cartel because each miner is independently selfish. For this reason we only consider a cartel if neither of them is individually selfish, i.e. $i$ and $j$ form a cartel if
$\hat{p}_{ij}<0.05$, while $\hat{p}_{i} \ge 0.05$ and $\hat{p}_{j} \ge 0.05$.


\subsection*{Address Clustering}
The blockchain protocol adopted by all the analysed cryptocurrencies except Ethereum allows users to have more than one address linked to their wallet, which might be used to hide the track of their transactions and balances. Thus, accurately clustering together the different addresses of a miner is very important to estimate the mining powers and detect miner behaviour. We applied three known methodologies to the different cryptocurrencies in our dataset \cite{meiklejohn2013fistful,androulaki2013evaluating,ermilov2017automatic} which are available from the blockchain analytics library BlockSci \cite{kalodner2020blocksci}:
\begin{itemize}
\item \textit{Heuristic 1 ($H_1$):} Multi-input Addresses\\
If two (or more) addresses are inputs to the same
transaction, they are controlled by the same user.
\item \textit{}{Heuristic 2 ($H_2$):} Optimal Change Address\\
If the amount of an output addresses is lower than any of the inputs used in the transaction, then it is reasonable to state that the output address is used for the transaction change and is controlled by the same user as the inputs.
\item \textit{Heuristic p ($H_p$):} Peeling chain \\
A transaction is considered to be in a peeling chain if it includes one input and two outputs, and both the previous and the following transaction follow this structure. It is reasonable to state that the outputs linking the peeling chain are change addresses.
\end{itemize}

Specifically, in the implementation, when using the first heuristic method (\textit{$H_1$}), different addresses used as inputs to one transaction are treated as being controlled by the same user. Then in \textit{$H_2$}, the identified so-called change addresses are treated as being controlled by the same user as the inputs. Finally in \textit{$H_p$}, the ``peeling chain'' structure is used as a different definition to identify change addresses.

\subsection*{Dataset}
Our empirical analysis focuses on five popular PoW-based cryptocurrencies which are Bitcoin (BTC), Litecoin (LTC), Monacoin (MONA), Ethereum (ETH) and Bitcoin Cash (BCH). Our dataset contains information of blocks from their launch to the end of 2020, including blocks' height, mined time, the tag of corresponding miners (miner address/ pool name). In detail, datasets of Monacoin and Ethereum only have the miner address of each block, while the datasets of other three cryptocurrencies already have the labeled name of mining pools \footnote{https://gz.blockchair.com/}. In this research, we define time windows for each cryptocurrency depending on the block time to ensure a similar amount of blocks in each detection interval. We present a summary description of the five datasets in Table \ref{tab:summary}, and further enumerate each crytocurrency for detailed introduction.    

\begin{table}[h]
\centering
\begin{tabularx}{0.8\textwidth}{rccrcc}

\hline 
Coin & Launch time & Block time & Height of blocks & Interval & Miners \\
\hline
Bitcoin & 2009-01-03 & 10 minutes & 663913  & monthly & Named pool  \\

Litecoin & 2011-10-07 & 2.5 minutes & 1974760 & weekly & Named pool \\
      
Monacoin & 2013-12-31 & 1.5 minutes & 2206670 & 5 days & Address\\

Ethereum & 2015-07-30 & 14 seconds & 11564743 & daily & Address\\

Bitcoin cash & 2017-08-01 & 10 minutes & 189735 & monthly & Named pool\\
\hline 
\end{tabularx}
\caption{\label{tab:summary}Summary description of datasets}
\end{table}

\begin{itemize}
    \item \textbf{Bitcoin} was started on 3 January 2009 when the internet persona Satoshi Nakamoto mined the first (so-called \textit{genesis})  block of the chain, known as the genesis block. Nowadays, this most famous digital asset has a rich and extensive ecosystem with a total market capitalisation of about 800 billions US dollars. About every 10 minutes, a new block is created and quickly published to all nodes, without requiring central oversight.
    \item \textbf{Litecoin} was a fork of the Bitcoin Core client released by Charlie Lee, a Google employee and former Engineering Director at Coinbase. The Litecoin network went live on 13 October 2011 differing primarily by having a decreased block generation time (2.5 minutes), increased maximum number of coins, different hashing algorithm, and a slightly modified GUI.
    \item \textbf{Monacoin} was a fork of Litecoin launched on 31 December 2013 by an anonymous person under the moniker of Mr. Watanabe. It bills itself as the first Japanese cryptocurrency and is predominantly used in Japan. Monacoin has an average block creation time of 1.5 minutes. Most notably, Monacoin was reported to have suffered from selfish mining attacks between May 13th and 15th in 2018 that caused roughly 90,000 dollars in damages \cite{saad2019countering}.
    \item \textbf{Ethereum} is the second largest cryptocurrency after Bitcoin, with currently over 300 billions US Dollars market capitalisation. Ethereum is the blockchain that issues Ether and was proposed in late 2013 by Vitalik Buterin, a cryptocurrency researcher and programmer, and the system went live on 30 July 2015 featuring smart contract functionality. The block time of Ethereum is around 14 seconds. In 2022, Ethereum will be moving from PoW to proof of stake (PoS) as part of its ``Ethereum 2.0'' upgrade.
    \item \textbf{Bitcoin Cash} was a hard fork of Bitcoin that seeks to add more transaction capacity to the network. On 1 August 2017, Amaury Séchet released the first Bitcoin Cash software implementation. As in Bitcoin, new blocks are on average generated every 10 minutes by using a difficulty adjustment algorithm (DAA). Bitcoin Cash also uses an Emergency Difficulty Adjustment (EDA) \cite{kwon2019bitcoin}, algorithm which has caused an instability in mining difficulty of the Bitcoin Cash system, resulting in Bitcoin Cash being thousands of blocks ahead of Bitcoin.

\end{itemize}

\bibliography{sample}


\section*{Acknowledgements}
S.-N.L. acknowledges funding from the China Scholarship Council (CSC) (No.201808310212). \\
C.C. acknowledges support from the Swiss National Science Foundation grant \#200021\_182659.

\section*{Author contributions statement}

C.J.T and S.-N.L. conceived the experiment(s). C.C and  S.-N.L. developed the methodology, S.-N.L. and C.J.T retrieved the data. S.-N.L.   performed the data analysis.  and C.C wrote a first draft.  All authors  completed and reviewed the final version of this paper. 

\section*{Competing interests}
The authors declare no competing interests.

\end{document}